\documentclass[10pt,letterpaper,twocolumn]{article} %% two column, final layout

\usepackage{ol2}
\usepackage[draft]{hyperref}
\usepackage{amsmath}

\begin{document}

\twocolumn[ %% activate for two-column option

\title{Coherent hard x-rays from attosecond \\
pulse train-assisted harmonic generation}

\author{Michael Klaiber$^{1,2}$, Karen Z. Hatsagortsyan$^1$, Carsten M\"uller$^{1}$ and  Christoph H. Keitel$^{1,*}$}

\address{
$^1$Max-Planck Institut f\"ur Kernphysik, Saupfercheckweg 1, D-69117 Heidelberg, Germany
\\
$^2$Theoretische Quantendynamik, Physikalisches Institut der Universit\"at, H.-Herder-Str. 3, D-79104 Freiburg, Germany
$^*$Corresponding author: keitel@mpi-hd.mpg.de 
}

\begin{abstract}
High-order harmonic generation from atomic systems is considered in the crossed fields of a relativistically strong infrared laser and a weak attosecond-pulse train of soft x-rays. Due to one-photon ionization by the x-ray pulse, the ionized electron obtains a starting momentum that compensates the relativistic drift which is induced by the laser magnetic field, and allows the electron to efficiently emit harmonic radiation upon recombination with the atomic core in the relativistic regime. In this way, short pulses of coherent hard x-rays of up to 40 keV energy and 10 as duration can be brought about.

\end{abstract}


 ] %% activate for two-column option

\noindent Coherent, short x-ray pulses are auspicious both for fundamental physics and applied science \cite{as}. One of the successful ways to produce coherent short-wavelength radiation is based on high-order harmonic generation (HHG) \cite{atto}. 
With this state-of-the-art technique, coherent x-ray photons of about 1 keV \cite{Seres} and short XUV pulses of about $100$ as \cite{Sansone} are generated via HHG in atomic gas jets. A further increase of the emitted photon energies cannot be achieved straightforwardly by an enhancement of the driving laser intensity as the interaction regime becomes relativistic and the returning electron misses the ionic core due to the relativistic drift that suppresses HHG \cite{review}. Various methods to counteract the relativistic drift have been proposed, which have either limited scope of applicability, or are challenging to realize \cite{drift,klaiber2}. On a different front, the attosecond laser technique is developing \cite{atto}. In particular, attosecond pulses superimposed onto an infrared laser field have recently been used to realize an attosecond streak camera \cite{streak_camera} and to control HHG spectra in the nonrelativistic regime \cite{carla}.

\begin{figure}[b]
  \begin{center}
    \includegraphics[width=0.3\textwidth,clip=true]{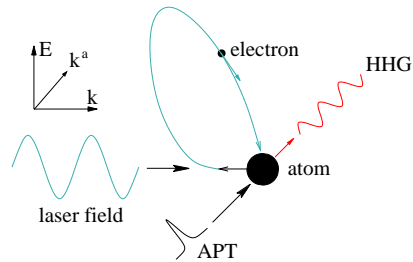}
    \caption{Scheme of the HHG process in the relativistic regime driven by crossed beams of an infrared laser field and an APT; $\mathbf{E}$, $\mathbf{k}$ and $\mathbf{k}^a$ are the laser polarization, laser propagation, and APT propagation directions.}
    \label{APTIR}
  \end{center}
\end{figure}

In this letter we investigate HHG in the relativistic regime which is driven by an infrared laser field crossed with an attosecond pulse train (APT) of soft x-ray radiation. In this case, the APT not only fixes the ionization time \cite{carla}, but also induces an initial electron momentum that can compensate the subsequent relativistic drift in the infrared laser field. This permits electron recollisions with the atomic core and efficient HHG in the relativistic domain. We employ a crossed-beam setup where the polarization direction of the APT is parallel to the propagation direction of the laser field. Then the electron emission probability in the direction along or opposite the laser propagation is the largest and, in the second case, the compensation of the relativistic drift is most efficient (see Fig.~\ref{APTIR}). We calculate the HHG yield in this setup in the moderately relativistic regime and show a large increase in the HHG yield as compared to the conventional case of a driving laser field only. 

Let us see how large the required energy $\Omega$ of the x-ray photons (equivalent to the carrier frequency of the APT) 
has to be in order to compensate the relativistic drift. The kinetic energy of the electron directly
after one-photon ionization by the x-ray pulse is
\begin{eqnarray}
  \varepsilon_0=\Omega-I_p,
  \label{e0}
\end{eqnarray}      
with the ionization potential $I_p$ (atomic units are used throughout). For the drift compensation, the initial energy has to be of the same order of magnitude as the drift energy: $\varepsilon_0 \approx p_d^2/2 \approx c^2\xi^4/32$, with the drift momentum $p_d\approx c\xi^2/4$ \cite{review}, the relativistic field strength parameter $\xi=E_0/c\omega$, the electric field of the laser $E_0$, the laser frequency $\omega$, and the speed of light $c$. The regime is relativistic when the electron drift distance is larger than the wave-packet spreading: $\sqrt{2I_p}(\xi^3/16)(c/\omega)>1$ \cite{review}. The intensity should be restricted, though, to prevent over-the-barrier ionization: $\xi < I_p^{3/2}/(4c\omega)$ \cite{review}. These conditions are satisfied in the relativistic tunneling regime with $\xi > 0.14$ and $I_p > 2.4$ a.u. We employ a moderately relativistic infrared laser field of $\xi \approx 0.33$ (ponderomotive energy $U_p=500$ a.u.) and $I_p=5.5$ a.u. (e.g., Be$^{2+}$ ions). Then the drift compensation demands soft x-rays of $\Omega \approx 9$ a.u. ($245$ eV) \cite{Tsakiris}. Under these conditions, a HHG cutoff $\omega_c\approx 3.17U_p$ of $43$ keV is attained.

We consider the interaction of an atomic system with the following superposition of a laser field and an APT.
The  laser field is propagating in $z$-direction and defined by the vector potential $\mathbf{A}(\eta)=\mathbf{e}_x (cE_0/\omega ) \cos \eta$, with $\eta=\omega (t-z/c)$ 
and the unit vector $\mathbf{e}_x$ in polarization direction. 
The pulses in the APT are Gaussian-shaped, propagate in $y$-direction, and have the electric field
$\mathbf{E}^{a}(y,t)=\mathbf{e}_zE^{a}_0g(y,t)\sin\Omega(t-y/c)$, with the unit vector $\mathbf{e}_z$ in polarization direction, 
the envelope $g(y,t)=\exp(-2(t-y/c-t_0)^2/\tau ^2)$, the time delay $t_0$ between the laser field and the APT, and the width $\tau$ of the pulses. The repetition rate of the attosecond pulses is twice the laser frequency.

The electron dynamics can be described by the Klein-Gordon equation in the single-active-electron approximation and the radiation gauge: 
\begin{eqnarray}
  \left(\partial^{\mu}\partial_{\mu}+c^2\right)\Psi({\rm x})=
  (V_L+V_X+V_{AI}+V_H)\Psi({\rm x}),
\end{eqnarray}
with the electron-ion interaction operator $V_{AI}=2iV/c^2\partial_{t}+V^2/c^2$, 
the atomic potential $V$, the time-space coordinate ${\rm x}=(ct,\mathbf{x})$, the electron-laser interaction operator $V_{L}=2i\mathbf{A}(\eta)\cdot\boldsymbol{\nabla}/c-\mathbf{A}(\eta)^2/c^2$,
the electron-APT interaction operator $V_{X}=2i\mathbf{A}^{a}(\eta)\cdot\boldsymbol{\nabla}/c-\mathbf{A}^{a}(\eta)^2/c^2$,
the APT vector potential $\mathbf{A}^{a}$, and the interaction operator between the electron and the quantized harmonic field $V_H({\rm x})=2\mathbf{A}_H({\rm x})/c\cdot (i\boldsymbol{\nabla}/c-\mathbf{A}_{tot}({\rm x})/c)$ (the $\mathbf{A}_H^2$ term is omitted as the harmonic field is a perturbation);
$\mathbf{A}_H({\rm x})=c\sqrt{2\pi/\omega_H}{\mathbf{e}}_H^*b^{\dagger}\exp(i\omega_Ht-i{\bf k}_H\cdot{\bf x})$ is the vector potential of the harmonic field in second quantization, 
with the unit vector ${\mathbf{e}}_H$ in polarization direction and the harmonic-photon creation operator $b^{\dagger}$, $\mathbf{A}_{tot}=\mathbf{A}+\mathbf{A}^{a}$, 
%$|l_H>$ are the states of the harmonic field with the number of photons $l$ ($l=0,1$),  
$\omega_H$ is the harmonic frequency, and $\mathbf{k}_H $ the wave vector. 
 
The differential HHG rate for the $n^{th}$ harmonic is  given by $dw_{n}/d\Omega=n(\omega/c)^3|M_n|^2$~\cite{klaiber2}, where the HHG amplitude in the strong-field approximation reads
\begin{eqnarray}
  M_{n}&=&
    -i \int
    d^4{\rm x}^{\prime}\int d^4{\rm x}^{\prime\prime} \left\{\Phi({\rm x}^{\prime})^* V_{H}({\rm x}^{\prime})
    \right. \nonumber \\
      &\times& \!\!\!\!\!  \left. G^V_L({\rm x}^{\prime},{\rm x}^{\prime\prime})
     \,\mathbf{x}^{\prime\prime}\cdot \left[\mathbf{E}({\rm x}^{\prime\prime})+\mathbf{E}^{a}({\rm x}^{\prime\prime})\right]\Phi({\rm x}^{\prime\prime})\right\},
     \label{mhhga}
\end{eqnarray}
with $\mathbf{E}=-\partial_t\mathbf{A}/c$. Here the APT can be treated in the dipole approximation since $E^{a}_0/\Omega c\ll 1$, and $c/\Omega \gg a_B$, with the Bohr radius $a_B$.
%$E^{a}_0\tau/c\ll 1$. 
Taking into account the shortness of the pulses in the APT ($\tau =4$ a.u.),
we further assume that the dynamics of the electron wave-packet after the ionization is governed by the laser field only. Therefore, the Klein-Gordon Green function in the total field can be replaced by the Volkov-Klein-Gordon Green function in the laser field $G^V_L({\rm x}^{\prime},{\rm x}^{\prime\prime})$~\cite{milo:01}. Note that the  wave function of the ground state $\Phi ({\rm x})$ is an eigenstate of the energy operator in the radiation gauge~\cite{klaiber1}.
The integral in Eq.~(\ref{mhhga}) is evaluated with the saddle-point method.

Two ionization scenarios are described by Eq.~(\ref{mhhga}). In the first scenario, the electron is both tunnel-ionized and propagated in the continuum by the laser field. In the other scenario, the process is initiated by the APT via single-photon ionization and the laser field takes over the electron in the continuum. In the relativistic regime, HHG due to the first scenario is heavily damped because of the relativistic drift. Therefore, we will consider the second scenario of the process. In this case, the saddle point equation reflects the energy conservation of Eq.(\ref{e0}) at the moment of ionization: $\varepsilon_{\mathbf{p}}(\eta^{\prime \prime})-c^2=\Omega-I_p$.

\begin{figure}[t]
  \begin{center}
    \includegraphics[width=0.3\textwidth,clip=true]{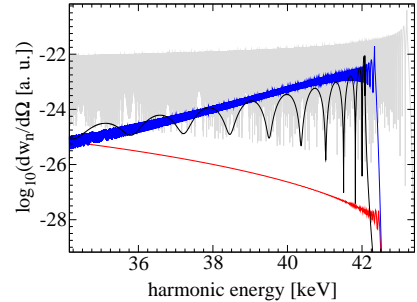}
    \caption{HHG spectra $dw_{n}/d\Omega$ via Eq.~(\ref{mhhga}) in a laser field $E_0=2.2$ a.u., $\omega=0.05$ a.u. for $I_p=5.5$ a.u.(Be$^{2+}$):
%Ar$^{9+}$ 
      (gray) dipole approximation, (red)
    Klein-Gordon equation, and additionally assisted by an APT $E^{a}_0=0.01$ a.u.,  carrier frequency $\Omega=9$ a.u., pulse length $\tau=4$ a.u., with a 
    phase delay of -1.2 rad and made of (black) Gaussian or (blue) delta pulses.}
    \label{APT_s}
  \end{center}
\end{figure}

In Fig.~\ref{APT_s} the HHG spectrum is displayed via Eq.~(\ref{mhhga}). 
A moderately relativistic infrared laser field of $2\times 10^{17}$ W/cm$^2$ ($\xi\approx 0.33$) and an APT of soft x-rays where each pulse has a duration of $100$ as and an energy of $300$ pJ is employed. 
Two cases are considered: 1) one-photon ionization by the APT
with ensuing propagation by the laser field and HHG in the APT propagation direction, and 2) tunneling ionization and propagation in a single laser field with HHG in the laser propagation direction. In the first case,
we use an APT with Gaussian pulses, or with delta pulses as in Ref.~\cite{carla}. For the second case, two calculations are presented: the first one is a fully relativistic calculation via the Klein-Gordon equation, whereas the second one treats the laser field in dipole approximation (DA). The latter serves as an indicator for ideal HHG via the tunneling-recombination mechanism without the relativistic drift. Due to the drift, in the sole laser wave the HHG signal is strongly suppressed by more than 6 orders of magnitude compared to the DA results. 
By superimposing a weak APT onto the infrared field, the HHG yield is enhanced by nearly 5 orders of magnitude and almost reaches the DA level. Moreover, the HHG rate can be increased proportionally to the APT intensity. Thus, the APT assistance enables recollisions with energies of 40 keV at significant rates. The description of the APT via delta-pulses is not justified for the parameters applied, since $\Omega\tau\ll 1$ \cite{darko}, but gives a qualitatively correct picture. Note that the different HHG emission directions considered in 1) and 2) ensure phase-matching between the radiation of different atoms.

\begin{figure}[t]
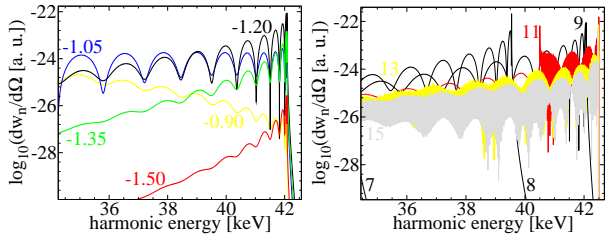

  \begin{center}
    \includegraphics[width=0.22\textwidth,clip=true]{klaiber_fig3a.eps}
    \includegraphics[width=0.22\textwidth,clip=true]{klaiber_fig3b.eps}
    \caption{HHG spectra $dw_{n}/d\Omega$ via Eq.~(\ref{mhhga}) in a laser field $E_0=2.2$ a.u., $\omega=0.05$ a.u. and an APT $E^{a}_0=0.01$ a.u., pulse duration $\tau=4$ a.u. for $I_p=5.5$ a.u.: (a) for different phase delays: -1.50, -1.35, -1.20, -1.15 and -0.90 rad, respectively; the carrier frequency is $\Omega=9$ a.u.;
(b) for different carrier frequencies: 7, 8, 9, 11, 13, and 15 a.u., respectively; the phase delay is -1.2 rad.
  }
    \label{APT_t}
  \end{center}
\end{figure}

We now investigate the stability of the HHG enhancement against variation of the APT parameters.
Firstly, we vary the phase delay between the two fields in Fig.~\ref{APT_t}a. Phase delay changes of about 0.15 rad have only a minor influence on the near-cutoff HHG rates, whereas changes of 0.3 rad reduce them significantly. This is due to the attosecond pulse duration of $\omega \tau=0.2$ rad: a phase delay variation
of this order or larger reduces the electric field in the pulse whenever trajectories leading to high final energies are launched.
This condition also determines the limited spatial area, where 
the APT assistance can strongly enhance the HHG yield.
Secondly, we vary the APT carrier frequency $\Omega$ at the optimal phase delay of the fields.
Fig.~\ref{APT_t}b shows that for small carrier frequencies the rate for high-energy harmonics is strongly reduced, since the energy supplied by the APT is too small to compensate the relativistic drift in the laser field. 
For larger values of the carrier frequency starting from $\Omega \approx 9$ a.u., efficient recombination with the atomic core becomes possible.
From the spectra shown, a slow decrease in the rates for higher carrier frequencies can be deduced. This is due to the scaling of the dipole-matrix element on the electron energy.
The optimal carrier frequency thus amounts to about 9 a.u.; higher values up to 11 a.u. slightly reduce the process efficiency.

A rough estimation, taking into account the spatial restriction of the phase matched emitters, shows that $100$ harmonic photons of $42$ keV photon energy can be emitted  within a $1$ keV spectral window at an ion density of $10^{16}$ cm$^{-3}$ (the electron excursion length is less than the interatomic distance), a laser pulse duration of $100$ fs, and an interaction volume of $10$ mm$^3$.

The HHG in the relativistic regime considered in this paper can also be used for generating ultra-short attosecond pulses via an appropriate filtering of harmonics. 
An analysis of the harmonic chirp shows that the optimal spectral  window is about 150 eV. Here, two trajectories are responsible for the HHG, which does not allow for a broad window. 
As a result, short pulses of 10 as duration can be produced.

In conclusion, we have shown that the newly emerging field of attosecond physics can
be applied for the benefit of modifying relativistic electron recollision dynamics in laser fields. 
With the proposed setup HHG with energies of 40 keV and significant rates as well as the generation of ultra-short pulses with a duration of 10 as is possible. Limiting factors for the respective macroscopic yields are the restricted areas which allow for phase-matching among the different ions and a high ion density.

This research was supported in part by the National Science Foundation under Grant No. PHY99-07949. We acknowledge the hospitality of the Kavli Institute for Theoretical Physics.

\end{document}